\documentclass[10pt,conference]{IEEEtran}
\IEEEoverridecommandlockouts

\usepackage{flushend}
\usepackage{verbatim}
\usepackage{color}
\usepackage{graphicx}
\usepackage{subfig}
\usepackage{wrapfig}
\usepackage{enumerate}
\usepackage[bookmarks=false]{hyperref}
\usepackage{comment}
\usepackage{amssymb}
\usepackage{amsmath}
\usepackage[ruled,vlined,linesnumbered,noresetcount]{algorithm2e}
\usepackage{caption}

\hypersetup{draft}  % edas does not like bookmarks generated by hyperref package

\usepackage[font=small,skip=2pt]{caption} % The skip reduces space between figure and caption.

\title{When Can You Trust Bitcoin? \\Value-Dependent Block Confirmation to\\ Determine Transaction Finality} 

\author{
Ethan Hicks(a), Joseph Oglio(b), Mikhail Nesterenko(a), and Gokarna Sharma(a)\\
(a) \textit{Department of Computer Science, Kent State University
Kent, OH 44242, USA}\\ and (b) \textit{School of Technology, Rio Grande University, Rio Grande, OH 45674, USA} \\
ehicks@kent.edu, joglio@rio.edu, mikhail@cs.kent.edu, sharma@cs.kent.edu}
\date{}

\begin{document}

\maketitle

\begin{abstract}
We study financial transaction confirmation finality in Bitcoin as a function of transaction amount and user risk tolerance. A transaction is recorded in a block on a blockchain.  
However, a transaction may be revoked due to a fork in the blockchain, the odds of which decrease over time but never reach zero.
Therefore, a transaction is considered confirmed if its block is sufficiently deep in the blockchain.  This depth is usually set empirically at some fixed number such as six blocks. We analyze forks under varying network delays in simulation and actual Bitcoin data. Based on this analysis, we establish a relationship between block depth and the probability of confirmation revocation due to a fork. We use prospect theory to relate transaction confirmation probability to transaction amount and user risk tolerance. 

% acceptable_probabiility = 0.01 * e^(-0.051 * 2.25 * dollars ^ 0.88)

\end{abstract}

\section{Introduction}
Bitcoin~\cite{nakamoto} is a distributed cryptocurrency based on blockchain. Bitcoin uses a competitive mechanism for transaction confirmation.  Clients submit transactions to the Bitcoin network of miners. Transactions are recorded on the blockchain. Each miner packages submitted transactions into a block. The block is cryptographically linked to the previous block forming a chain. To earn rewards, miners compete to add these blocks to the blockchain. Due to the network delay or, possibly, malicious miner activity, multiple blocks may be linked to the same parent block. This is a \emph{fork}. Miners may continue to add blocks to the prongs of the fork. However, only the blocks on the longer prong are confirmed. Thus, it is in the interest of miners to add blocks to the longest chain. Eventually, the miners move to mine on the longer prong and the fork is resolved. However, the transactions on the shorter prong are revoked. 

Due to network delay, a client may not be able to see which prong is extended in a timely manner. Thus, it may not be clear if a particular transaction is confirmed or revoked due to a fork. Moreover, in modern Bitcoin, most mining is carried out by mining pools rather than individual miners. A \emph{mining pool} is a group of miners that work jointly to increase their combined mining rate and share the proceeds. The mining rate of some pools exceeds $30\%$  of the mining rate of the entire network. Such a high mining rate may enable an attacker to revoke undesirable transactions with ease~\cite{eyal2018majority, fehnker2018twenty}.

% this number is from 2013 and likely has changed... I don't know if we should mention this
More than $1.5\%$ of blocks in Bitcoin are forked~\cite{decker2013information} and all of these need to be resolved causing the revocation of many transactions. The possibility of transaction revocation decreases as more blocks are added on top of a transaction as both prongs of a fork are being extended at roughly the same time is just as likely as a fork. However, this probability never reaches zero. Thus, the decision whether to consider the confirmation final lies with the user and is related to the user's willingness to risk losing the transaction money. This means that if a vendor accepts a transaction that is later revoked, the vendor may lose the associated funds with no recourse.

Usually, it is recommended to wait a certain amount of time or for a certain number of blocks to be added to the blockchain on top of the transaction for the confirmation to be considered final. It is also recommended to wait longer for large value transactions. However, longer wait decreases the 
usefulness of the cryptocurrency. In particular, it is difficult to use Bitcoin for small-value transactions~\cite{bamert2013have}. Hence, transaction finality wait time needs to be selected with care. Yet, we are not aware of any study that formally relates this finality to the transaction amount or user risk tolerance. Nor is it known whether traditional waiting recommendations were checked against user risk tolerance. 

\ \\
\textbf{Contribution.} In this paper, we use Prospect Theory~\cite{kai1979prospect,tversky1992advances} to derive a confirmation–risk curve. This allows practitioners, vendors, and crypto traders to individually select their confirmation finality time depending on their risk tolerance, transaction amount, and network fork formation conditions. 
We calibrate our findings using QUANTAS~\cite{quantas} abstract distributed algorithm simulator and actual Bitcoin data.

\section{Bitcoin and Risk Modeling}

\noindent\textbf{Bitcoin terminology.}
A \emph{flipped block} is a block on the discarded prong of the fork. A transaction in a flipped block is revoked.
The \emph{confirmation depth} of a transaction is the number of blocks appended after its block. A greater depth reduces the risk of revocation but increases delay in servicing the transaction. A \emph{switch} is the event of a miner receiving a longer prong of a fork from other miners and thus switching to mining on this longer prong. For a client that observes the switching miner, such switch revokes the transactions in the blocks of the shorter prong.  \emph{Switch depth} is the depth of the shorter prong that contains revoked transactions.

%Thegreater the confirmation depth of a transaction, the lower the probability that a switch occurs and the transaction is revoked. 

\begin{figure}[htbp]
  \centering
  \includegraphics[width=0.49\textwidth]{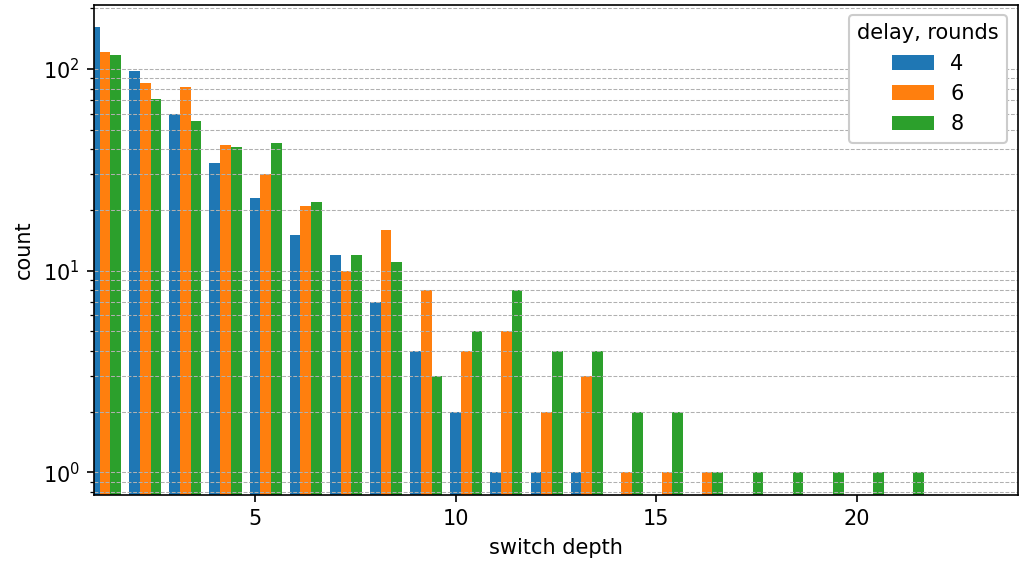}
  \caption{Simulation. Number of switches of particular lengths depending on the network delay. }
  \label{fig:simSwitches}
\end{figure}

\begin{figure}[htbp]
  \centering
  \includegraphics[width=0.49\textwidth]{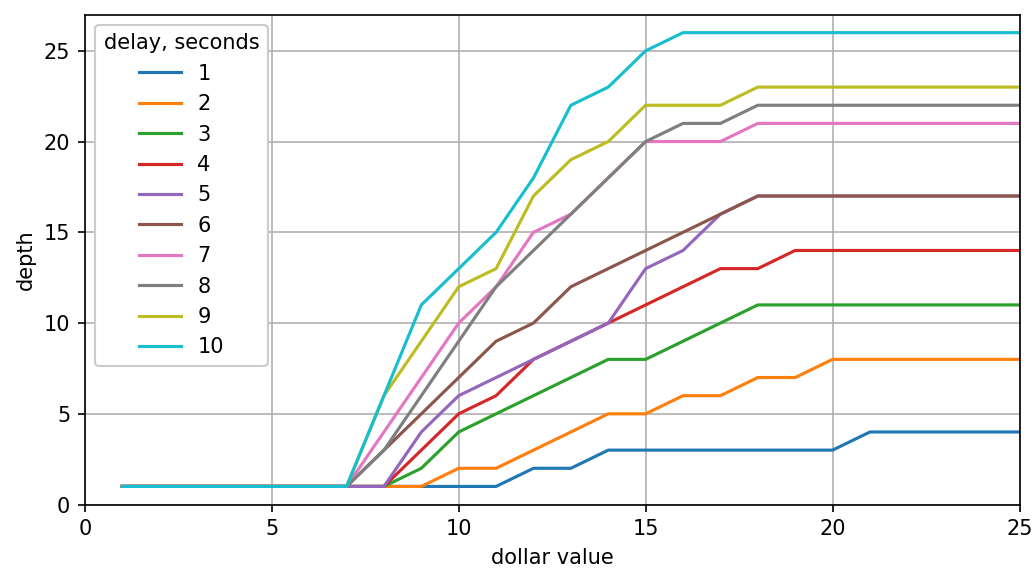}
  \caption{Simulation. Minimum confirmation depth vs dollar value for various network delays.}
  \label{fig:simDepth}
\end{figure}

\ \\
\noindent
\textbf{Applying Prospect Theory to Bitcoin transactions.}
Prospect Theory models user preferences under uncertainty. Let us introduce two functions: loss probability and loss.
The loss probability function can be used to determine the likelihood of transaction revocation and loss of funds. And the loss function can be used to measure the acceptable likelihood of losing an amount of funds i.e. the amount of risk someone is willing to take on. We define a loss function $L$ for transaction revocation as a function of transaction value $v$:
\[ L(v) = -\lambda \cdot (-v)^{\beta} \]
where  $\lambda > 1$ is a loss aversion coefficient that models a human tendency to weigh loses more heavily than gains;  the exponent $0 < \beta < 1$ models a diminishing sensitivity to higher losses. The following parameter values $\lambda = 2.25$ and $\beta = 0.88$  are shown to accurately measure human aversion to risk~\cite{tversky1992advances}.

We define the loss threshold probability $LT(v)$ as the user's transaction revocation risk tolerance. In our case, above this threshold the user thinks that the transaction revocation risk is too great for the given value of the transaction to consider its confirmation final and will instead wait for more blocks to be added to reduce the risk. 

This loss threshold is computed as: 
\[ LT(v) = e^{L(v)/c} \]
where $c$ is set such that $LT(1) = 0.5$. This means that a transaction that is worth $\$1$ is revoked with $50\%$ probability. This value was chosen as a transparent behavioral anchor but could be adjusted as necessary: a single dollar is often considered consequential enough to trigger deliberation but small enough that many people are indifferent to either losing it or keeping it. For transaction confirmation to be final, the probability of transaction revocation $P_{\text{rev}}$ at depth $d$ should not exceed this threshold:
\[ P_{\text{rev}}(d) \leq LT(v) \]
$P_{\text{rev}}$  depends on the depth of a transaction. We determine its value through both simulation and actual blockchain data.
The transaction depth determines the time a client has to wait for a transaction to finalize. Hence, we try to minimize this depth. The \emph{minimum confirmation depth} is the shallowest depth that satisfies the above formula.

\begin{table}[htbp]
  \centering
  \begin{tabular}{l r}
    \hline
    \textbf{mining pool} & \textbf{blocks mined}\\
    \hline
    foundrydigital.com & 299\\
    antpool.com        & 189\\
    viabtc.com         & 129\\
    f2pool.com         & 100\\
    spiderpool.com     & 68\\
    mara.com           & 54\\
    luxor.tech         & 38\\
    secpool.com        & 30\\
    binance.com        & 19\\
    braiins.com        & 17\\
    sbicrypto.com      & 16\\
    ntminerpool.com    & 10\\
    cloverpool.com     &  7\\
    ultimuspool.com    &  6\\
    ocean.xyz          &  6\\
    poolin.com         &  4\\
    whitebit.com       &  3\\
    nicehash.com       &  1\\
    SoloCKPool         &  1\\
    Unknown1           &  1\\
    Unknown2           &  1\\
    Unknown3           &  1\\
    \hline
  \end{tabular}
  \caption{Distribution of blocks mined in the most recent 1,000 Bitcoin blocks by pool.}
  \label{tab:block-dist}
\end{table}

\begin{figure}[htbp]
  \centering
  \includegraphics[width=0.49\textwidth]{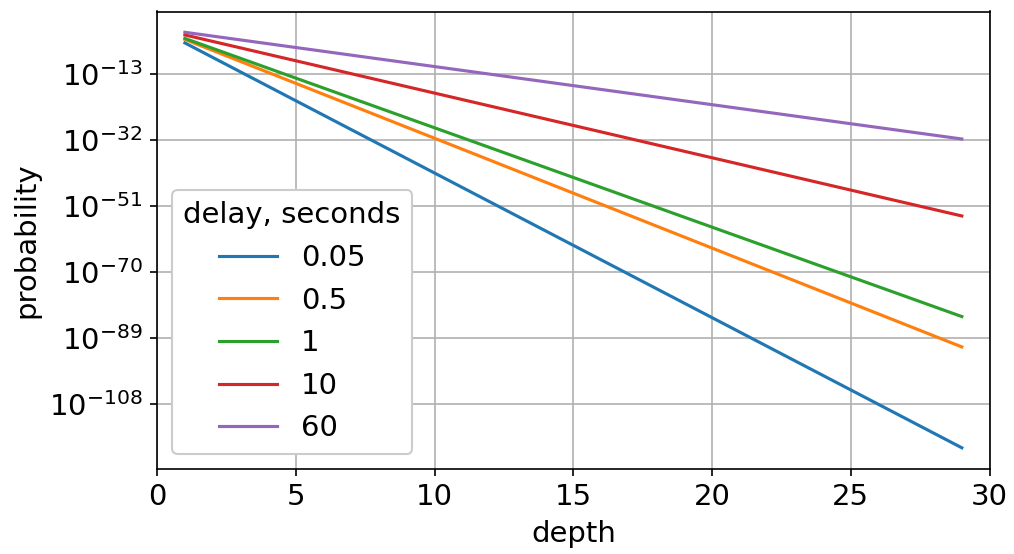}
  \caption{Actual Bitcoin Data. Probability of  revocation vs. depth.}
  \label{fig:actualProbability}
\end{figure}

\begin{figure}[htbp]
  \centering
  \includegraphics[width=0.49\textwidth]{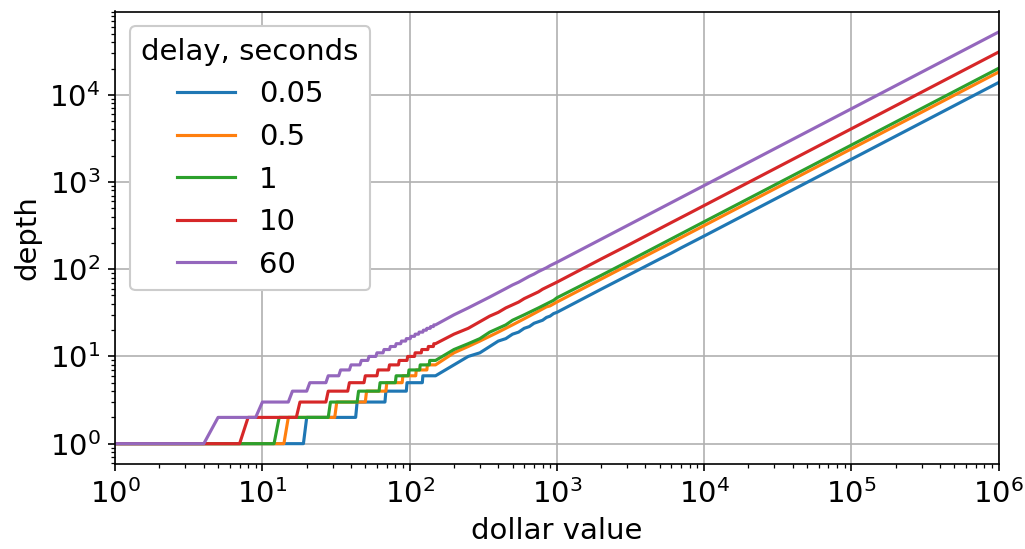}
  \caption{Actual Bitcoin Data. Minimum confirmation depth vs dollar value for various network delays.}
  \label{fig:actualDepth}
\end{figure}

\section{Simulation Setup and Results}

\noindent
\textbf{Simulation setup.}
We simulate a Bitcoin network in QUANTAS. QUANTAS is an abstract distributed algorithms simulator. The simulated computation proceeds in a sequence of rounds where each peer receives previously sent messages, does local processing, then sends messages to be received by other peers in the subsequent rounds. The round when the message is received is determined by the message delay. We simulate a network of $100$ miners and study it under various message delays. For each round and each miner, we record the number of switches of a particular length.
For each data point, we run $10$ computations of  $1000$ rounds each and average the obtained results across these computations. 

\ \\
\textbf{Simulation results.}
Figure \ref{fig:simSwitches} shows the distribution of switch depths under network delays of $4$, $6$, and $8$ rounds. This demonstrates that the number of deeper switches increases with the network delay. Put another way, as the network delay increases, so does the probability of a transaction being revoked at higher depths. 
Figure \ref{fig:simDepth} plots the relationship between transaction value and minimum confirmation depth under network delays ranging from $1$ to $10$ rounds. The data demonstrates how minimum confirmation depth requirements scale with both transaction value and network delay.
For example, with a delay of $1$, the transaction of $\$15$ has to wait $3$ blocks above it to be finalized. Alternatively, a delay of $10$ requires the transaction of $\$15$ to wait for $25$ blocks to finalize. 

\begin{comment}
With a delay of $1$ the minimum confirmation depth of $4$ blocks is reached at approximately $v \approx \$21$.

In contrast, under significantly degraded network conditions, the maximum necessary confirmation depth increases to $26$ blocks at only $v \approx \$16$. This substantial difference illustrates the profound impact of network delay on security requirements.

The observed behavior aligns with Prospect Theory loss thresholds, where confirmation depth requirements are inversely proportional to the probability of block revocation for small transaction values. As transaction value increases alongside network delay, the necessary confirmation depth grows correspondingly. This relationship reflects the fundamental principle that higher network instability, induced by increased propagation delays, necessitates an increase in confirmation depth to ensure transaction finality and prevent potential reversals due to forks.
\end{comment}

% \textbf{Metrics and Analysis}
% Key findings include:
% \begin{itemize}
%   \item \textbf{Blockchain Length:} Longer under higher delay due to increased block contention.
%   \item \textbf{Flipped Blocks:} Increases with delay, reflecting higher fork frequency.
%   \item \textbf{Switches:} Number of times longest chain changes.
%   \item \textbf{Finality Probability:} Probability a block is retained grows with depth.
% \end{itemize}

% Block height 900338 to 899338

\section{Analysis of Actual Bitcoin Data}
We examined data from the last $1000$ blocks mined on the main chain of Bitcoin~\cite{MiningPoolStatsBitcoin}. We determined the relative block mining rates of the mining pools that mined these blocks, see Table~\ref{tab:block-dist}. We considered network delays ranging from $.05$ to $60$ seconds. It has been shown that it takes on average $6.5 s$ for half the network to receive a block, but $5 \%$ of nodes take $40 s$. However, nearly all nodes have a delay of less than $60 s$ which makes it a reasonable worst-case value~\cite{decker2013information}. On the basis of the pool mining rates, we computed the probability of a pool mining a block while another mined block was in transition due to the network delay. That is, we determined the probability of revocation at a depth of one. Probabilities of deeper block revocation are computed as a product of the depth one revocation probability.

Figure \ref{fig:actualProbability} shows the probability that a block of a certain depth is revoked depending on the network delay.
% The network delays range from $0.05$ to $60$ seconds. 
% we do not know and should not speculate, MN
%The curves follow an exponential decay with the depth. 
As the block depth increases, its revocation probability declines. This decline is shallower with greater network delay. 
Figure \ref{fig:actualDepth} plots the relationship between transaction dollar value and minimum confirmation depth. It is similar to Figure~\ref{fig:simDepth}. In actual data, a $\$100$ transaction requires approximately $6$ blocks at a network delay of $1$ second. Note that our simulation is more aggressive in the fork probability and we explore the blockchain behavior at far greater network delays than can be observed in actual Bitcoin operation.  
The graphs in Figures~\ref{fig:simDepth} and~\ref{fig:actualDepth} are of particular use to practitioners since they show the estimated wait time until a transaction is finalized depending on the network conditions, transaction value, and user tolerance to risk. 

\section{Conclusion and Future Work} 

We quantified Bitcoin transaction finality through simulation and actual analysis enabling practitioners, vendors, and crypto traders to make informed financial decisions rather than relying on ad hoc recommendations.  We anticipate multiple future research directions based on this work. We may use the actual data to calibrate our simulation to better reflect Bitcoin operation. Observe that our research may be applied in a straightforward manner to other popular competitive blockchains such as Ethereum Classic, Bitcoin Cash, Bitcoin SV etc.~\cite{ethereum, cash2019bitcoin}.

In this work, we assume that network conditions are fixed. However, real-time network monitoring and delay estimation may be used to adjust transaction finality recommendations. We do not provide recommendations as to how to use transaction finality. However, we envision automated ``smart wallets'' that use this information to report to users when their transaction is finalized.  

%\clearpage

\bibliographystyle{IEEEtran}
\bibliography{references}
\end{document}